\documentclass[conference, a4paper]{IEEEtran}
\usepackage[utf8]{inputenc}
\usepackage[english]{babel}
\usepackage{amsmath}
\usepackage{amssymb}
\usepackage{cmap}
\usepackage{cite}
\usepackage{setspace}
\usepackage{amsmath}
\usepackage{mathtools}
\usepackage{hyperref}
\usepackage{siunitx}
\usepackage{subcaption}
\usepackage{tikz}
\usetikzlibrary{arrows}
\usetikzlibrary{patterns}

\usepackage{graphicx}
\graphicspath{{images/}}

\newcommand{\bbP}{\mathbb{P}}
\IEEEoverridecommandlockouts

\newcommand\copyrighttext{%
\footnotesize \textcopyright \enspace 2020 IEEE. Personal use of this material is permitted. Permission from IEEE must be obtained for all other uses, in any current or future media, including reprinting/republishing this material for advertising or promotional purposes, creating new collective works, for resale or redistribution to servers or lists, or reuse of any copyrighted component of this work in other works. DOI: \href{https://doi.org/10.1109/ICUMT51630.2020.9222409}{10.1109/ICUMT51630.2020.9222409}
}
\newcommand\copyrightnotice{%
\begin{tikzpicture}[remember picture,overlay]
\node[anchor=south] at (current page.south) {\fbox{\parbox{\dimexpr\textwidth-\fboxsep-\fboxrule\relax}{\copyrighttext}}};
\end{tikzpicture}%
}

\begin{document}
\title{Tuning Channel Access to Enable \\ Real-Time Applications in Wi-Fi 7 \thanks{The research was done at MIEM NRU HSE and supported by the Russian Science Foundation (agreement No 18-19-00580).}
}
\author{\IEEEauthorblockN{
Dmitry Bankov\IEEEauthorrefmark{1}\IEEEauthorrefmark{2}\IEEEauthorrefmark{3},
Kirill Chemrov\IEEEauthorrefmark{2}\IEEEauthorrefmark{3},
Evgeny Khorov\IEEEauthorrefmark{1}\IEEEauthorrefmark{2}\IEEEauthorrefmark{3}
}
\IEEEauthorrefmark{1}National Research University Higher School of Economics, Moscow, Russia\\
\IEEEauthorblockA{\IEEEauthorrefmark{2}Institute for Information Transmission Problems, Russian Academy of Sciences, Moscow, Russia\\
\IEEEauthorrefmark{3}Moscow Institute of Physics and Technology, Moscow, Russia\\
Email: \{bankov, chemrov, khorov\}@wireless.iitp.ru}
}

\maketitle
\copyrightnotice

\begin{abstract}
	Real-Time Applications (RTA) are among the most important use cases for future Wi-Fi 7, defined by the IEEE 802.11be standard.
	This paper studies two backward-compatible channel access approaches to satisfy the strict quality of service (QoS) requirements of RTA on the transmission latency and packet loss rate that have been considered in the 802.11be Task Group.
	The first approach is based on limiting the transmission duration of non-RTA frames in the network.
	The second approach is based on preliminary channel access to ensure the timely delivery of RTA frames.
	With the developed mathematical model of these approaches, it is shown that both of them can satisfy the RTA QoS requirements.
	At the same time, the preliminary channel access  provides up to 60\% higher efficiency  of the channel usage by the non-RTA traffic in scenarios with very strict RTA QoS requirements or with low intensity of the RTA traffic.
\end{abstract}


\section{Introduction}{\label{sec:intro}}
Real-Time Applications (RTA) are steadily becoming an inherent part of our lives.
RTA are present in such scenarios, as cloud gaming, video streaming, virtual and augmented reality, distance learning, and industrial automation.
Devices that support RTA become more and more available, and their traffic grows every year.
RTA have very strict quality of service (QoS) requirements: typical RTA demand very low latency ($\leq $\SI{10}{\ms}) and very high reliability of communications (packet loss rate, PLR $\leq 10^{-5}$)~\cite{discussion_target_presentation, usecases_presentation}.
At the same time, in many scenarios, it is either inconvenient or impossible to use wired network technologies, which is a motivation to develop approaches for RTA service in wireless networks.

The IEEE 802 LAN/MAN Working Group considers RTA as a very important use case for the next generation of Wi-Fi.
In 2018 the RTA Topic Interest Group (TIG) was created, which classified the RTA-related scenarios and proposed methods to provide the QoS for RTA in Wi-Fi~\cite{rta_tig}. 
RTA support has become one of the major goals of the next main Wi-Fi standard amendment: the IEEE 802.11be, a.k.a. Wi-Fi 7 \cite{khorov2020current}.

It is not easy to guarantee a low delay in Wi-Fi, mostly because Wi-Fi stations (STAs) use a variant of Carrier Sense Multiple Access with Collision Avoidance (CSMA/CA) to access the channel.
CSMA/CA does not allow a STA to interrupt the transmission of another STA, so even a high-priority STA with an urgent frame has to wait until some other STA frees the channel, and the waiting time can reach $\approx \SI{5}{\ms}$.
Additional delays can be caused by the frame collisions that occur when several STAs access the channel at the same time.

A possible standard way to decrease the RTA frame delay in Wi-Fi networks is to limit the time by which the STAs can occupy the channel, and thus to decrease the time that a STA has to wait until the channel becomes idle.
Still, the overhead induced by the channel access procedure and the frame headers decrease the channel usage efficiency for other STAs in the network.
Another approach that we propose and study in this paper is the preliminary channel access (PCA).
Its idea is that in case of periodic and/or predictable RTA traffic, the STA can obtain the channel access in advance and reserve the channel by sending a Request-to-Send (RTS) frame to have the transmission opportunity (TXOP) when the RTA frame is generated.
Such an approach also decreases the channel usage efficiency for the other STAs.
In the paper, we compare these approaches to determine their efficiency and to find out the conditions when one approach or the other one should be used for better performance.	
To achieve this goal, we develop mathematical models of networks using the considered approaches, state and solve the optimization problem to maximize the channel usage efficiency, provided that the delay quantile does not exceed a given constraint.

The rest of the paper is organized as follows.
In Section~\ref{sec:related}, we review the prior arts on RTA service in Wi-Fi.
In Section~\ref{sec:methods}, we provide details on the Wi-Fi channel access and the studied approaches.
Section~\ref{sec:problem} describes the scenario and the problem statement.
In Section~\ref{sec:model}, we develop a mathematical model of the RTA service approaches.
Section~\ref{sec:numerical} contains the numerical results.
Conclusion is given in Section~\ref{sec:outro}.

\section{Related Works}{\label{sec:related}}
Solutions for RTA service in Wi-Fi are studied both in IEEE 802.11 task groups and in scientific papers.
A promising approach for RTA service is based on orthogonal frequency division multiple access (OFDMA) and has been proposed in the RTA TIG and studied in~\cite{avdotin2019enabling, avdotin2020resource}.
OFDMA was introduced in the 802.11ax amendment and will be further developed in 802.11be.
With OFDMA, a Wi-Fi Access Point (AP) can divide the frequency band into resource units and allocate them to STAs so that they can transmit their data in parallel.
By controlling the STA transmissions, the AP can improve the efficiency of channel usage, and several algorithms have been proposed to combine the deterministic and random channel access in order to provide low delays for RTA.

Another way to prioritize the RTA traffic is to use an additional radio interface to signal about the RTA traffic presence~\cite{bankov2019enabling}.
Having received such a signal with additional radio interface, all the devices that transmit the non-RTA frames must free the channel at once.

The RTA QoS can be improved using the Multi-Link transmissions~\cite{rodriguezmulti}, an approach that is currently in the center of attention of the 802.11be task group.
With Multi-Link, a STA can contend for the medium in several frequency channels at once and transmit its data in the first free channel.
A similar approach based on double Wi-Fi interfaces has been considered \cite{cena2017experimental} outside the 802.11be amendment.

The approaches described so far require significant changes to the Wi-Fi standard and additional PHY level functionality, which can hinder their introduction into real devices.
However, much simpler and easy-to-implement solutions exist, e.g., \cite{genc2019wi} propose to introduce a new access category (AC) for RTA traffic and to set its channel access parameters to guarantee that the RTA traffic is served before the non-RTA one.
Another solution is to allow the STA to start the channel access procedure in advance if it expects a forthcoming RTA frame~\cite{abouelseoudreducing}.

In this paper, we develop the last two ideas.
We consider a separate AC for RTA traffic, and in addition to~\cite{genc2019wi}, we set its channel access parameters in such a way that the channel usage efficiency by non-RTA traffic does not suffer too much.
We also consider PCA, but in addition to~\cite{abouelseoudreducing}, we expand it with sending an RTS frame.
The proposed methods are quite simple and can be easily implemented in real devices.

\section{Channel Access for RTA}{\label{sec:methods}}

\subsection{Channel Access in Wi-Fi}\label{ssec:edca}
Modern Wi-Fi STAs access the channel according to the Enhanced Distributed Channel Access (EDCA).
Every STA maintains a queue of frames waiting for transmission.
If a STA generates a frame when its queue is empty, it listens to the channel and, if it is idle, transmits the frame at once.
If the channel is busy, the STA initializes a backoff counter with an integer random number distributed uniformly in the $[0, CW_r - 1]$ interval, where $r$ is the retry counter (initially zero), and $CW_r$ is the contention window defined as
\begin{equation}
CW_r = 
\begin{cases}
CW_{min}, &	r=0;\\
\min\{2CW_{r-1}, CW_{max}\}, & r \in(0, RL].
\end{cases}
\end{equation}
Here $CW_{min}$ and $CW_{max}$ are the minimal and maximal contention window, respectively, and $RL$ is the retry limit.

The backoff counter is frozen as long as the channel is busy, and is unfrozen if the channel is idle for the Arbitration Interframe Space ($AIFS$).
An unfrozen backoff counter decrements by one every empty slot interval $T_e$.
The STA transmits its frame when the backoff counter reaches zero.

Short Interframe Space ($SIFS)$ after a successful frame reception, the receiving STA should reply with and acknowledgment frame (ACK).
Having received an ACK, the STA sets its retry counter to zero and starts processing the next frame, if any.
If the transmitter STA does not detect an ACK within the $AckTimeout$ interval, it considers the transmission as failed, increases the retry counter, and performs a new transmission attempt.
The STA discards the frame if it makes $RL$ unsuccessful transmission attempts.

To provide QoS for different types of traffic, EDCA distinguishes between several access categories (ACs).
Every AC has its queue and backoff function. 
Different ACs have different AIFS values equal to $SIFS + T_e \cdot AIFSN$, where $AIFSN$ is an integer from 2 to 15.
Moreover, when a STA wins the channel access, it gains the channel for a time interval, called the transmission opportunity (TXOP) and bounded by the TXOP limit.
Every AC has its values of $CW_{min}$, $CW_{max}$, $AIFSN$, and TXOP limit, which are set by the AP and are the same for all STAs in the network.

\subsection{RTS/CTS Procedure}\label{ssec:rts}
\begin{figure}[tb]
	\centering
	\includegraphics[width=0.9\linewidth]{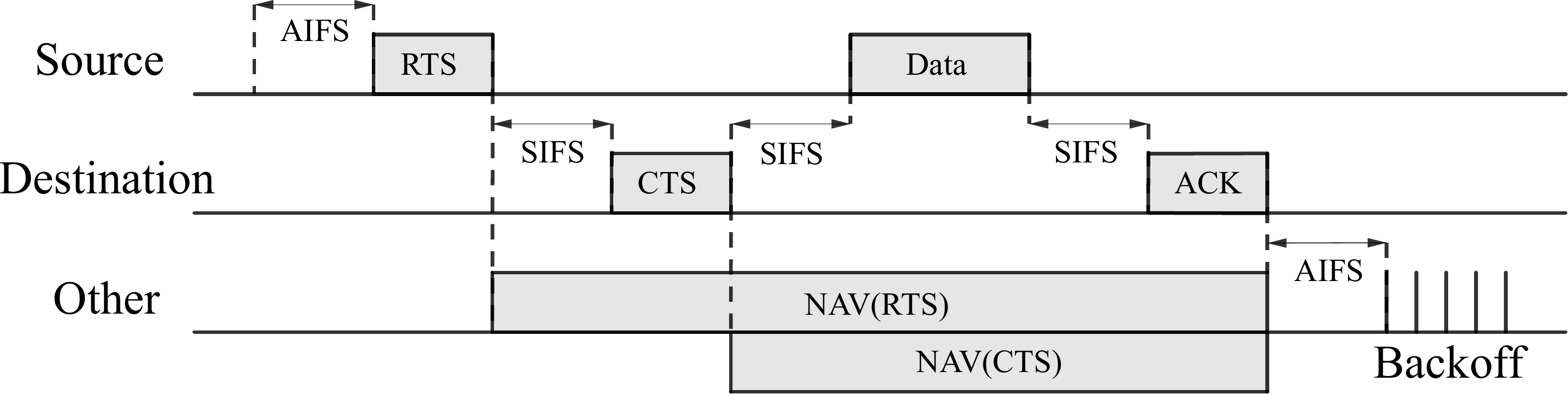}
	\caption{\label{fig:rts} The RTS/CTS Procedure}
\end{figure}

To increase the channel usage efficiency for long frames, Wi-Fi networks use the RTS/CTS procedure (see Fig.~\ref{fig:rts}).
With this procedure, a STA transmits an RTS frame, the receiving STA responds with a Clear-to-Send (CTS) frame, and only after receiving a CTS, the STA transmits the data frame.
Even though the RTS and CTS transmission introduces additional overhead, this procedure improves the channel usage because the RTS and CTS are very short, and a collision of RTS frames lasts much less than a collision of long data frames.

The RTS and CTS, as well as most Wi-Fi frames, contain a Duration/ID field, which specifies how long the channel will be busy after the end of the frame.
Having received a frame with such a field, the STA sets up its network allocation vector (NAV): a counter of virtually busy channel time, and may not access the channel as long as NAV is non-zero.
A STA can increase the NAV duration if the Duration/ID field of a received frame is higher than NAV, but may not decrease it.
The only exception is the receipt of a CF-end frame, which means that the STA should zero its NAV.

\subsection{RTA Service Approaches}\label{ssec:methods}
The described above functionality can be used to provide reliable low latency service for RTA.
First, we assign a separate AC for RTA traffic, and set $AIFSN_{RTA}$ value less than the $AIFSN$ for other ACs.
Let $\Delta_{AC}$ be the difference between the smallest $AIFSN$ for non-RTA traffic and $AIFSN_{RTA}$.
To ensure that RTA traffic has a priority over non-RTA traffic, we set its contention windows as follows:
\begin{equation}
CW_{RTA} \triangleq CW_{max}^{RTA} = CW_{min}^{RTA} \leqslant \Delta_{AC}.
\end{equation}
The RTA frames thus will always win the contention for channel access with non-RTA frames, but still the RTA frames may experience delay caused by waiting for ongoing transmissions to end.
To decrease this delay, we consider two approaches.
The first approach is to set a small TXOP limit for non-RTA ACs, and thus to restrict the waiting time for the RTA frames \cite{tsn_presentation}.

An alternative approach is to capture the channel by sending RTA before the frame arrives in the transmission queue \cite{abouelseoudreducing}.
With this approach, $T_b$ before the expected RTA frame arrival (see Fig.~\ref{fig:rts_delay}), the STA generates an RTS.
If the RTS is successfully delivered, the STA can transmit an RTA frame without delay once the frame is generated.
It should be noted that the 802.11 standard does not consider the RTS transmissions without a present data frame, so we need to add such a capability to the 802.11be amendment. 

Both approaches consume the channel resources, and it is a question, how to configure them to satisfy the RTA QoS requirements and to maximize the network performance.

\section{Scenario and Problem Statement}{\label{sec:problem}}
We consider a Wi-Fi network consisting of an AP, $N$ legacy (non-RTA) STAs, and one RTA STA.
All legacy STAs generate a saturated stream of data frames with a non-RTA AC with parameters $CW_{min}$, $CW_{max}$, $AIFSN$ and TXOP limit $= T_s$ set by the AP.
We assume that the application of RTA STA periodically generates single data frames, but due to the implementation issues, such as the unpredictable operation system delays or the clock drift, the real frame arrival time is random.
The expected RTA frame generation time constitutes a periodic process with a period of $T_{period}$, but the real arrival time $t_a$ is distributed normally around the expected one with a standard deviation $\sigma$.
The RTA AC has parameters $CW_{RTA}$ and $AIFSN_{RTA} = AIFSN - \Delta_{AC}$.

The legacy STAs use the RTS/CTS procedure to transmit their frames, and the data frame size is such that the transmission duration equals the TXOP limit, including the inter-frame spaces, RTS, CTS, data frame and ACK transmission.

The RTA frames have a fixed size and the following QoS requirements: the frame transmission delay should be less than $D_{max}$, and PLR should be less than $PLR_{QoS}$.

We consider two approaches described in Section~\ref{ssec:methods}.
In the first approach we configure the TXOP limit depending on $D_{max}$ to achieve the required $PLR_{QoS}$.
In the second approach, we take into account $D_{max}$ and $\sigma$ and configure the time offset $T_b$ between the expected RTA frame generation and the RTS transmission start.
We assume that after a successful RTS transmission, the STA can keep the channel until the data frame arrival using the NAV mechanism or by sending some busy channel signal.
After a successful data frame transmission, the STA ends the transmission using a CF-end frame.
If the data frame is generated before the RTS transmission start, the STA transmits the data frame instead.

In the described scenario, we state the problem to find the optimal TXOP limit $= T_s$ and $T_b$ parameters for the two approaches and to determine which approach provides better performance depending on the traffic parameters and QoS requirements.
We consider two utility functions.
The first one is the average channel usage efficiency $E$: the portion of channel time occupied by successful non-RTA frame transmissions.
The second one is the $1 - PLR_{QoS}$ quantile of delay $Q$, i.e., such a delay value, that with probability $1 - PLR_{QoS}$ an RTA frame is transmitted within this delay.
We state the following optimization task for the approach without PCA:
\begin{equation}\label{eq:opt_simple}
\begin{split}
\max_{T_s}  \quad E_{simple}(T_s), \quad 
s.t. \quad Q_{simple}(T_s) \leqslant D_{max},
\end{split}
\end{equation}
and with PCA:
\begin{equation}\label{eq:opt_reserve}
\begin{split}
\max_{T_b, T_s} E_{PCA}(T_b, T_s),  \quad 
s.t. \  Q_{PCA}(T_b, T_s) \leqslant D_{max}.
\end{split}
\end{equation}

\section{Mathematical Model}{\label{sec:model}}
The mathematical model consists of two main parts.
First, we consider a rarely-transmitting RTA STA as a small perturbation in comparison with several saturated legacy STAs, and describe the legacy STAs behavior as if there was no RTA STA.
Second, we consider the RTA STA and find the delay distribution and its influence on the channel usage efficiency.	

\subsection{Legacy STAs}
We describe the saturated legacy STAs similarly to~\cite{bianchi2000performance, vishnevsky2002802}.
We assume that the STAs can hear each others' transmissions and therefore count their backoff synchronously.
We divide the time into variable-duration slots, their bounds corresponding to the backoff countdowns by the STAs. 
Slots can be empty, successful, and collision.
A slot is empty if no STAs try to transmit their frames during the slot.
Its duration is $T_e$.
A successful slot occurs when only one STA transmits its frame.
The successful slot duration equals $T_s + AIFS$, where $T_s = T_{RTS} + SIFS + T_{CTS} + SIFS + T_{data} + SIFS + T_{ACK}$.
Here $T_{RTS}$, $T_{CTS}$, $T_{data}$, and $T_{ACK}$ are the RTS, CTS, data frame and ACK durations, respectively.
A collision slot is a slot during which more than one STAs try to transmit.
Its duration equals $T_c + AIFS$, where $T_c = T_{RTS} + AckTimeout$.

Let us choose a legacy STA.
Let $\tau$ be the probability of the STA transmission in a slot, and let $p$ be the probability of STA transmission attempt resulting in a collision.
If the STA tries to transmit in a slot, the probability of the remaining $N - 1$ STAs not to transmit in the slot equals $(1-\tau)^{N-1}$, so 
\begin{equation}\label{eq:p}
p = 1 - (1-\tau)^{N-1}.
\end{equation}

As in \cite{vishnevsky2002802}, we find $\tau$ as the average number of transmissions divided by the average number of slots counted per frame: 
\begin{equation}\label{eq:tau}
\tau = \left.\left(\sum\limits_{r=0}^{RL}p^r\right)\middle/ \left(\sum\limits_{r=0}^{RL}\frac{CW_r-1}{2}p^r\right)\right.,
\end{equation}
where $p^r$ is the probability of STA making the transmission attempt $r+1$, and $\frac{CW_r-1}{2}$ is the average backoff.

Solving the system of equations \eqref{eq:p} and \eqref{eq:tau}, we find  $p$ and $\tau$, and use them to find the probabilities of an empty slot: $P_e = (1-\tau)^N$, successful slot: $P_s = N\tau(1-\tau)^{N-1}$, and collision slot: $P_c = 1 - P_e - P_s$.

\subsection{RTA Frame Delay}
The STA generates an RTS $T_b$ before the expected arrival of the RTA frame.
If the channel is empty, which happens with probability $P_{te}$, the STA transmits the RTS at once.
Otherwise, the STA waits for the channel to become idle and starts the backoff countdown before the RTS transmission.
The channel is busy with a successful transmission with the probability $P_{ts}$ and with a collision with a probability $P_{tc}$.

\begin{figure}[tb]
	\centering
	\includegraphics[width=0.6\linewidth]{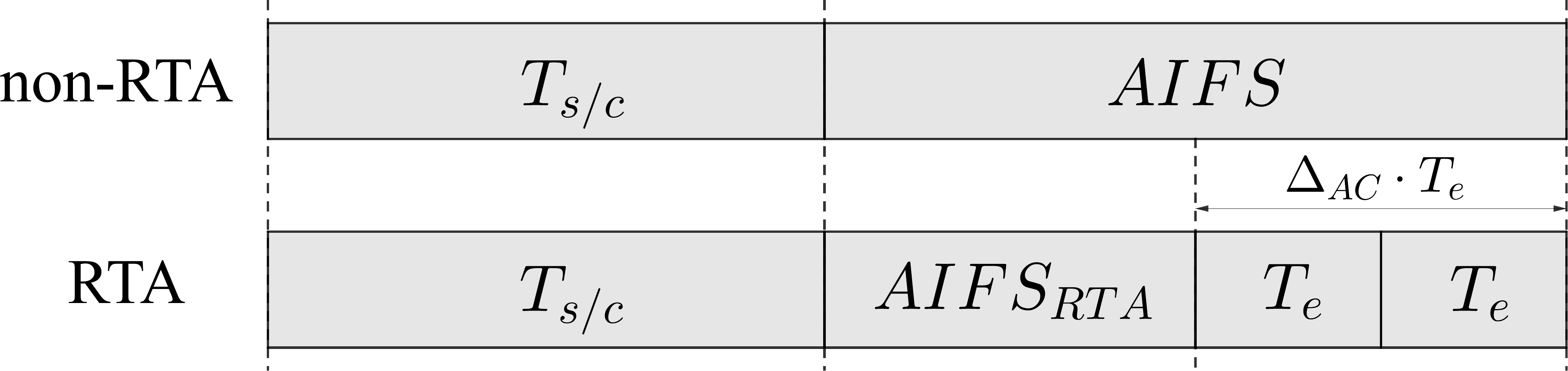}
	\caption{\label{fig:slots} Comparison of RTA and non-RTA Slots}
\end{figure}

We define the delay $D$ as the time from the data frame arrival until the end of its transmission.
This value depends on the type of the slot, during which the RTA is generated.
At the same time, due to the difference in $AIFS$ values, the RTA STA observes more empty slots and shorter successful and collision slots in comparison with the legacy STAs (see Fig. \ref{fig:slots}).
The average number of empty slots for RTA STAs is bigger by $\Delta_{AC}\cdot(P_s+P_c)$, while the successful and collision slot durations consist of $T_s$, $T_c$ and $AIFS_{RTA}$.
Taking these values into account, we find the portion of time, when the channel is empty, occupied by a successful transmission or a collision from the point of view of the RTA STA:
\begin{equation*}
\begin{aligned}
P_{te} &= \frac{P_eT_e + (P_s+P_c)\cdot \Delta_{AC}\cdot T_e}{P_eT_e + P_s(T_s+AIFS) + P_c(T_c+AIFS)}, \\
P_{ts} &= \frac{P_s(T_s + AIFS_{RTA})}{P_eT_e + P_s(T_s+AIFS) + P_c(T_c+AIFS)}, \\
P_{tc} &= \frac{P_c(T_c + AIFS_{RTA})}{P_eT_e + P_s(T_s+AIFS) + P_c(T_c+AIFS)}.
\end{aligned}
\end{equation*}

\begin{figure}[tb]
	\centering
	\includegraphics[width=\linewidth]{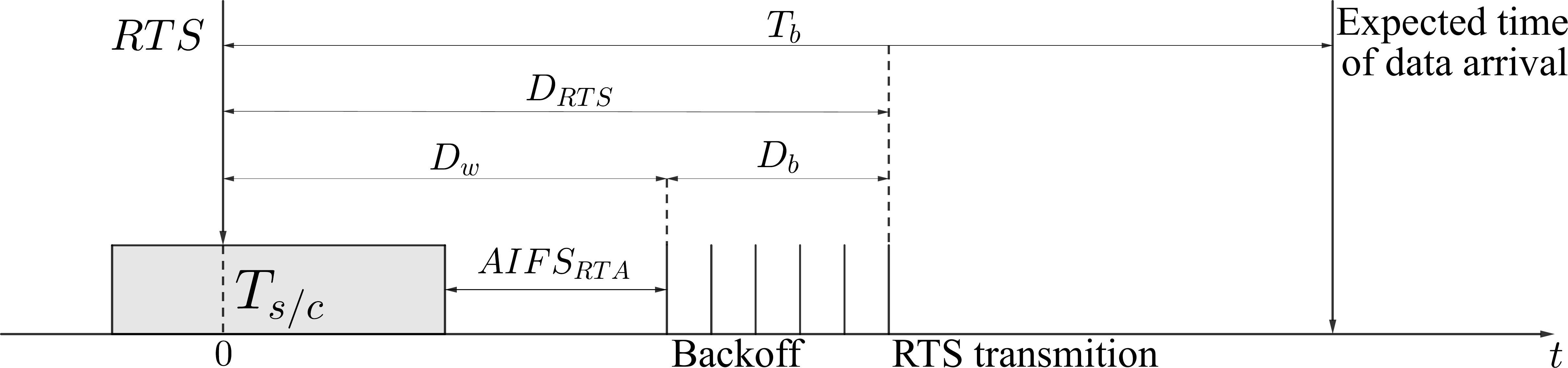}
	\caption{\label{fig:rts_delay} Components of RTS Delay}
\end{figure}

Thus, with probability $P_{te}$, the RTS is sent at once.
Otherwise, the RTS frame experiences a random delay $D_{RTS}$, which includes the waiting time for the channel to become idle $D_w$ (including $AIFS_{RTA}$), and the backoff countdown $D_{b}$~(see Fig.~\ref{fig:rts_delay}): $D_{RTS} = D_w + D_b$.

The RTS can arrive with equal probability in any time within a slot, so $D_w$ is distributed uniformly with a CDF:
\begin{equation}\label{eq:D_w}
F_{w,s/c}(t) = 
\begin{cases}
0, &	t<AIFS_{RTA};\\
\frac{t}{T_{s/c}}, & AIFS_{RTA} \leqslant t < T_{s/c} + AIFS_{RTA};\\
1, & t \geqslant T_{s/c} + AIFS_{RTA},
\end{cases}
\end{equation}
where $s / c$ denotes whether the random value describes a successful or a collision slot.
We combine this CDF with $D_b$: an integer value distributed  uniformly from 0 to $CW_{RTA} - 1$, and find the RTS delay CDF:
\begin{equation*}
\resizebox{\linewidth}{!}{$F_{RTS,s/c}(t)=\bbP(D_w + D_b \leqslant t) = \sum\limits_{i=0}^{CW_{RTA}-1}\frac{F_{w,s/c}(t - i\cdot T_e)}{CW_{RTA}}.$}
\end{equation*}

Taking into account three types of slots we obtain:
\begin{equation}\label{eq:D_rts}
F_{RTS}(t) = P_{te} + P_{ts}\cdot F_{RTS,s}(t) +  P_{tc}\cdot F_{RTS,c}(t),
\end{equation}
if $t \geqslant 0$, and $0$ otherwise.

Let the RTS be generated at time $0$.
The data frame arrival time $t_a$ has the following PDF:
\begin{equation}\label{eq:f_ta}
f_{t_a}(t) = \frac{1}{\sigma\sqrt{2\pi}}\exp\left(-\frac{(t-T_b)^2}{2\sigma^2}\right).
\end{equation}

If an RTS is transmitted before the data frame arrival, the channel is reserved by the RTA STA, and the delay equals the transmission time $T_{SR}$, which includes the RTA frame, $SIFS$, and ACK.
Otherwise, the frame can arrive before the RTS generation ($t_a < 0$).
In this case, the RTA frame delay is distributed as $F_{RTS}(t - T_{SR})$.

\begin{figure}[tb]
	\centering
	\includegraphics[width=\linewidth]{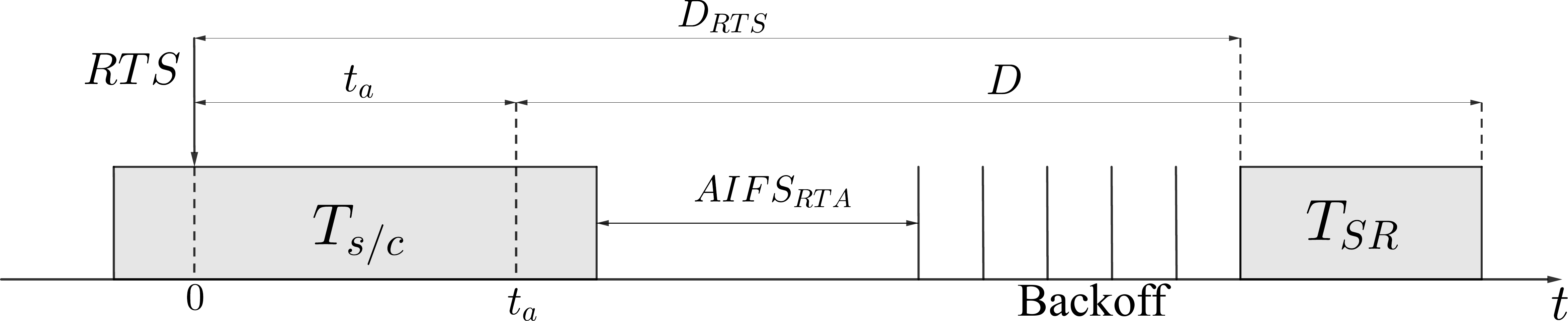}
	\caption{\label{fig:data_delay} Components of Data Transmission Delay for $t_a>0$}
\end{figure}

The RTA frame can arrive after the RTS generation but before its transmission ($0 < t_a < D_{RTS}$).
In such a case, the delay consists of the rest of the $D_{RTS}$ and the RTA frame transmission time: $D = D_{RTS} - t_a + T_{SR}$ (see Fig.~\ref{fig:data_delay}).

Combining the described cases, we find the delay CDF for PCA:
\begin{equation}
\begin{split}
F_{D,PCA}(x) &= F_{RTS}(x-T_{SR})\int\limits_{-\infty}^{0}f_{t_a}(t)dt +\\
+& \int\limits_{0}^{\infty}f_{t_a}(t)F_{RTS}(t+x-T_{SR})dt.
\end{split}
\end{equation}

In the approach without PCA, the STA transmits the data frame instead of RTS, so the delay consists only of the channel access time, its CDF being
\begin{equation}
F_{D,simple}(x) = F_{RTS}(x-T_{SR}).
\end{equation}

\subsection{Channel Usage Efficiency}
Without the RTA STA, we define the channel usage efficiency as the portion of time when the channel is occupied by a successful data transmission:
\begin{equation}\label{eq:Eff_wo_rta}
\resizebox{0.88\linewidth}{!}{$E_{\text{w/o RTA}} = \frac{P_s\cdot T_{payload}}{P_eT_e + P_s(T_s+AIFS) + P_c(T_c+AIFS)}$,}
\end{equation}
where $T_{payload}(T_s) = T_s - (T_{RTS} + SIFS + T_{CTS} + T_{header} + SIFS + T_{ACK})$ is the payload transmission time and $T_{header}$ is the frame header duration.

If we add an RTA STA, we need to take into account that once in $T_{period}$ it occupies the channel for $T_{RTA}$ time:
\begin{equation}\label{eq:Eff}
E = \left(1-\frac{ T_{RTA} }{T_{period} }\right) \cdot E_{\text{w/o RTA}}.
\end{equation}

Without PCA, the RTA STA occupies the channel only for the data frame duration and $AIFS$: $T_{RTA} = T_{SR} + AIFS_{RTA}$.

PCA increases $T_{RTA}$ value by $t_a - D_{RTS}$, and we find the average increase value using~(\ref{eq:f_ta}) and $f_{RTS} = \frac{dF_{RTS}(t)}{dt}$.
\begin{equation*}
\resizebox{\linewidth}{!}{$\int\limits_{0}^{\infty}\int\limits_{0}^{x}(x-t)\cdot f_{t_a}(x)\cdot f_{RTS}(t)dtdx =
\int\limits_{0}^{\infty} f_{t_a}(x) \int\limits_{0}^{x}  F_{RTS}(t)dtdx.$}
\end{equation*}

For PCA, we add the duration of RTS, CTS, and CF-end.
In total, we obtain:
\begin{equation*}
\begin{split}
&\resizebox{\linewidth}{!}{$T_{RTA} = T_{SR} + AIFS_{RTA} + \int\limits_{0}^{\infty} f_{t_a}(x) \int\limits_{0}^{x}  F_{RTS}(t)dtdx+$} \\
&\resizebox{\linewidth}{!}{$+ (T_{RTS}+2 \cdot SIFS+T_{CTS}+T_{CFend}) \int\limits_{0}^{\infty} f_{t_a}(x)F_{RTS}(x)dx$.}
\end{split}
\end{equation*}

Given $T_{RTA}$, we find $E$ with~\eqref{eq:Eff}.

\subsection{Optimization Problem}
\label{sec:optimization}
To solve the optimization tasks \eqref{eq:opt_simple} and \eqref{eq:opt_reserve}, we find the delay quantile $Q$ that corresponds to the successful transmission probability $1 - PLR_{QoS}$.
For that we solve the equations:
\begin{equation*}
\begin{split}
F_{D,simple}(Q_{simple}(T_s)) &= 1-PLR_{QoS},\\
F_{D,PCA}(Q_{PCA}(T_b, T_s)) &= 1-PLR_{QoS}.
\end{split}
\end{equation*}
These equations define the maximal delay $D_{max}$ as a function of the $T_s$ and $T_b$ parameters.

Let us fix all the parameters except $T_b$ and $T_s$.
Equation~\eqref{eq:Eff_wo_rta} can be reduced to the following form:
\begin{equation*}
\resizebox{\linewidth}{!}{$E_{\text{w/o RTA}}(T_s) = 1 - \frac{P_eT_e + P_s(AIFS+const) + P_c(T_c+AIFS)}{P_eT_e + P_s(T_s+AIFS) + P_c(T_c+AIFS)}.$}
\end{equation*}
Now we see that to solve the optimization task~\eqref{eq:opt_simple} for the approach without PCA, we should set $T_s$ to the highest value that does not break the limitation on the delay quantile.
The maximal $T_s$ is found as a solution of the equation:
\begin{equation}
Q_{simple}(T_s) = D_{max}.
\end{equation}

For PCA, we also maximize $T_s$.
At the same time, if we increase $T_b$, we decrease the delay quantile, but increase the channel time consumed by the RTA transmission.
To solve the optimization task, we fix the maximal possible $T_s$ value and find the minimal $T_b$ which satisfies the delay constraint:
$
Q_{PCA}(T_b) = D_{max}.
$

\section{Numerical Results}
\label{sec:numerical}
We model a Wi-Fi network described in Section~\ref{sec:problem} with a simulation and the developed analytical model.
The modeling parameters are listed in Table~\ref{tab:param}.
We consider that all STAs use the modulation and coding scheme MCS0.

\begin{figure}[t]
	\centering
	\begin{subfigure}{\linewidth}
		\includegraphics[width=\textwidth]{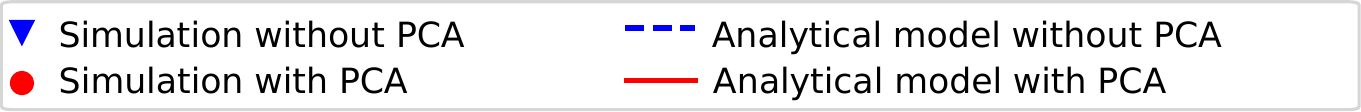}
	\end{subfigure}
	\begin{subfigure}{0.49\linewidth}
		\includegraphics[width=\textwidth]{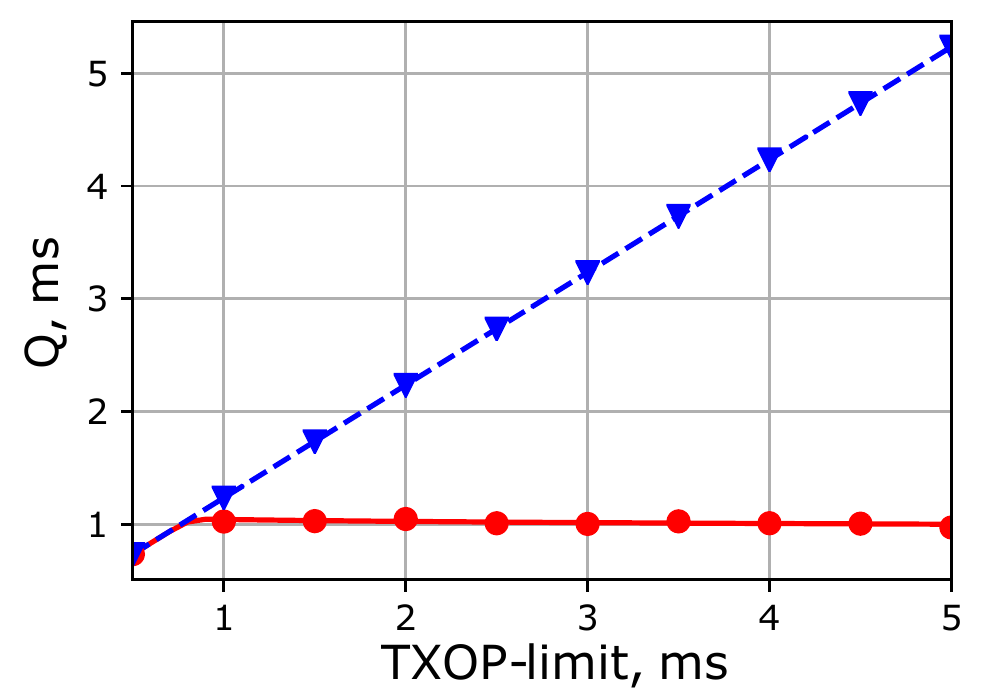}
		\caption{\label{fig:Q_s1_d1}$\sigma=\SI{0.1}{\ms}$, $D_{max}=\SI{1}{\ms}$}
	\end{subfigure}
	\begin{subfigure}{0.49\linewidth}
		\includegraphics[width=\textwidth]{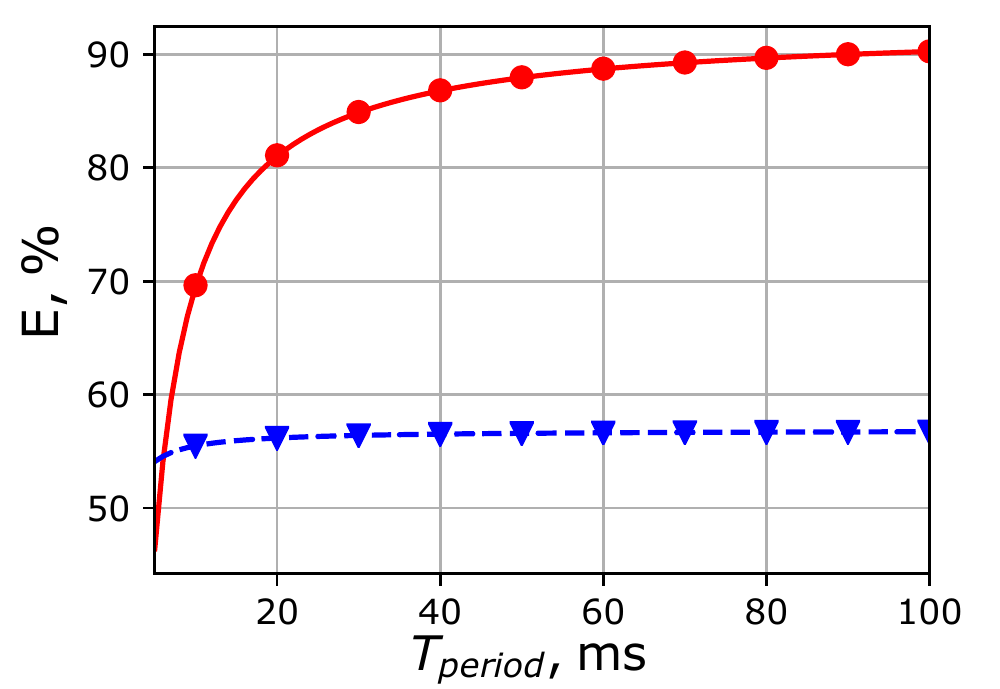}
		\caption{\label{fig:E_s1_d1}$\sigma=\SI{0.1}{\ms}$, $D_{max}=\SI{1}{\ms}$}
	\end{subfigure}
	\begin{subfigure}{0.49\linewidth}
		\includegraphics[width=\textwidth]{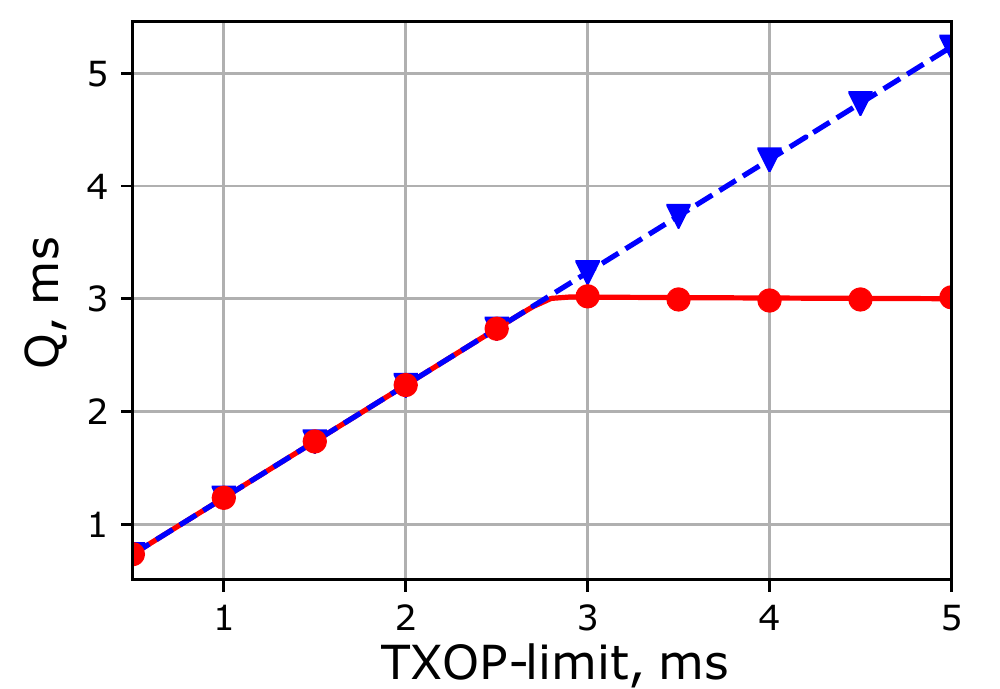}
		\caption{\label{fig:Q_s1_d3}$\sigma=\SI{0.1}{\ms}$, $D_{max}=\SI{3}{\ms}$}
	\end{subfigure}
	\begin{subfigure}{0.49\linewidth}
		\includegraphics[width=\textwidth]{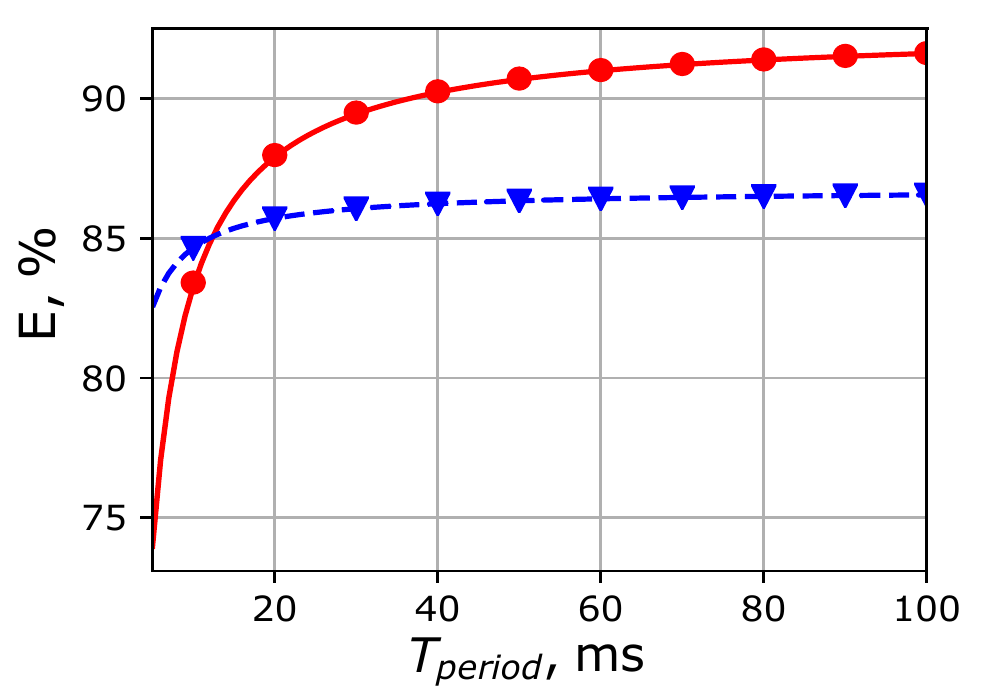}
		\caption{\label{fig:E_s1_d3}$\sigma=\SI{0.1}{\ms}$, $D_{max}=\SI{3}{\ms}$}
	\end{subfigure}
	\begin{subfigure}{0.49\linewidth}
		\includegraphics[width=\textwidth]{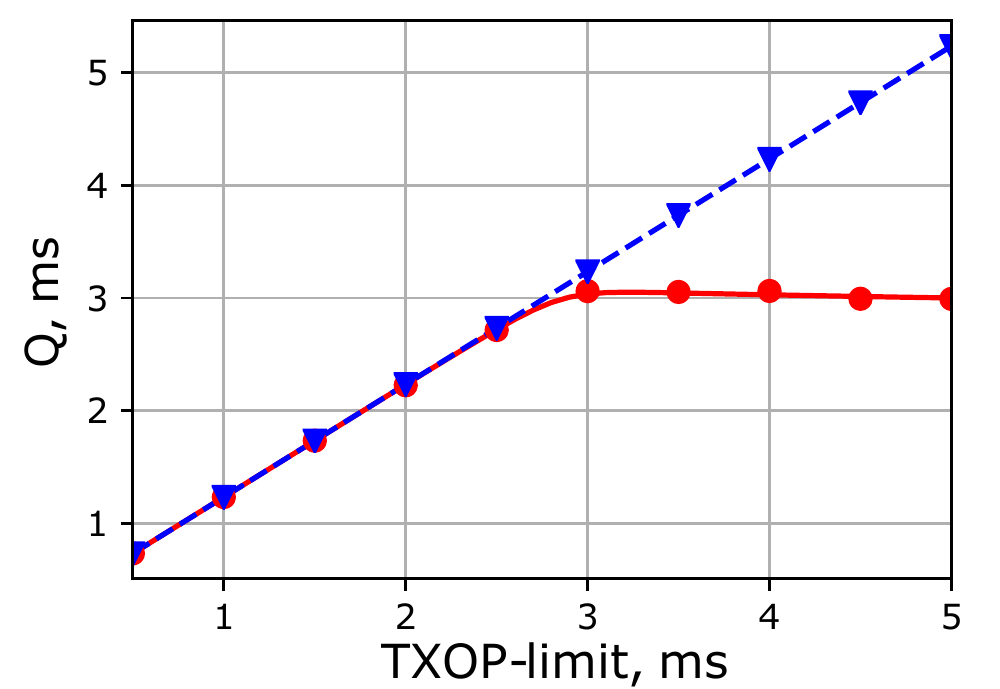}
		\caption{\label{fig:Q_s5_d3}$\sigma=\SI{0.5}{\ms}$, $D_{max}=\SI{3}{\ms}$}
	\end{subfigure}
	\begin{subfigure}{0.49\linewidth}
		\includegraphics[width=\textwidth]{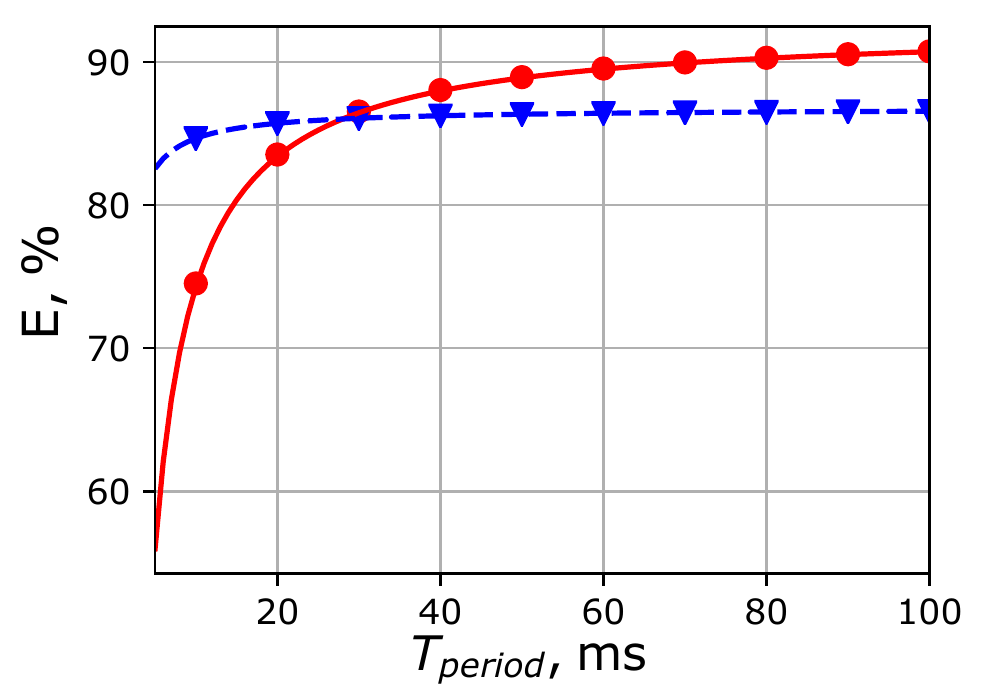}
		\caption{\label{fig:E_s5_d3}$\sigma=\SI{0.5}{\ms}$, $D_{max}=\SI{3}{\ms}$}
	\end{subfigure}
	\begin{subfigure}{0.49\linewidth}
		\includegraphics[width=\textwidth]{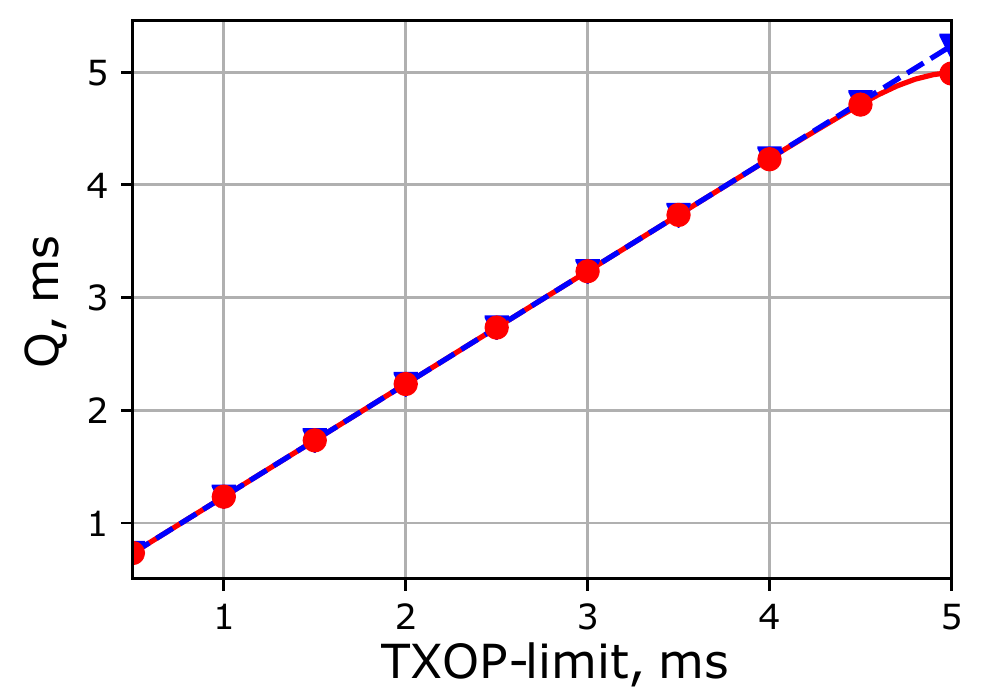}
		\caption{\label{fig:Q_s5_d5}$\sigma=\SI{0.5}{\ms}$, $D_{max}=\SI{5}{\ms}$}
	\end{subfigure}
	\begin{subfigure}{0.49\linewidth}
		\includegraphics[width=\textwidth]{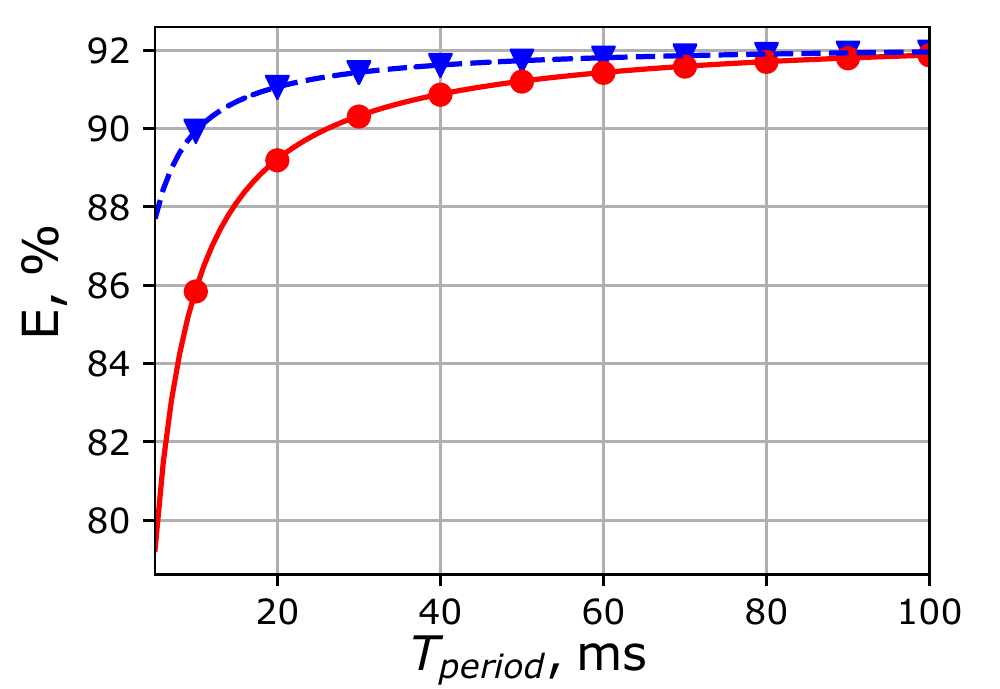}
		\caption{\label{fig:E_s5_d5}$\sigma=\SI{0.5}{\ms}$, $D_{max}=\SI{5}{\ms}$}
	\end{subfigure}
	\caption{\label{fig:plots}Dependency of delay quantile on TXOP limit and dependency of the channel usage efficiency on $T_{period}$}
\end{figure}

Fig.~\ref{fig:plots} shows the dependency of the delay quantile $Q$ on the TXOP limit and the dependency of the average efficiency of channel usage $E$ on the RTA frame period $T_{period}$.
The numerical results are obtained with simulation and analytical model for several values of $\sigma$ and maximal delay $D_{max}$.

For the reservation approach, we set the TXOP limit to \SI{5}{\ms} to achieve a high channel usage efficiency.
As shown in Fig.~\ref{fig:plots}, the delay quantile does not exceed $D_{max}$ for smaller TXOP limit.
Such a result is caused by the fact that the corresponding optimal $T_b$ value is found as $T_b = TXOP~limit - const$, where $const$ is optimized according to $\sigma$ and $D_{max}$.

\begin{table}[t]
	\caption{Modeling Parameters}\label{tab:param}
	\begin{center}
		\begin{tabular}{c|c||c|c}
			\hline
			\textbf{Parameter} & \textbf{Value} & \textbf{Parameter} & \textbf{Value} \\ \hline
			$T_e$			& \SI{9}{\us}		& $SIFS$		& \SI{16}{\us}		\\ \hline
			$T_{SR}$		& \SI{191.2}{\us}	& $T_{header}$	& \SI{40}{\us}		\\ \hline
			$AckTimeout$	& \SI{53}{\us}		& $PLR_{QoS}$	& $10^{-5}$			\\ \hline
			$AIFSN$			& 4					& $CW_{RTA} = \Delta_{AC}$ & 2		\\ \hline
			$CW_{min}$		& 16				& $CW_{max}$	& 1024				\\ \hline
			$RL$			& 7					& $N$			& 10				\\ \hline
		\end{tabular}
	\end{center}
\end{table}

From Fig.~\ref{fig:plots}, we see that for a very small RTA frame generation period, decreasing TXOP limit yields better channel usage than PCA while providing the required delay quantile.
PCA becomes more efficient for the RTA frame periodicity above $T^{*}_{period}$, which can be found by solving the equation
\begin{equation}
E_{simple}(T^{*}_{period}) = E_{reserve}(T^{*}_{period}).
\end{equation}

Fig.~\ref{fig:E_s1_d1} and Fig.~\ref{fig:E_s1_d3} show that if we increase the maximal delay from \SI{1}{\ms} to \SI{3}{\ms}, the $T^{*}_{period}$ also grows from \SI{6}{\ms} to \SI{12.5}{\ms}.
At the same time, according Fig.~\ref{fig:E_s1_d3} and Fig.~\ref{fig:E_s5_d3}, the increase in the standard deviation of RTA frame arrival time from \SI{0.1}{\ms} to \SI{0.5}{\ms} also increases the $T^{*}_{period}$ to \SI{28}{\ms}.
Thus the range of RTA periodicity for which PCA is better decreases with the growth of $\sigma$ and $D_{max}$.

The maximal gain in the channel usage efficiency also depends on $\sigma$ and $D_{max}$.
For $\sigma=\SI{0.1}{\ms}$ and $D_{max}=\SI{1}{\ms}$ (Fig.~\ref{fig:E_s1_d1}) the gain is almost $60\%$.
If we increase $D_{max}$ to \SI{3}{\ms} (Fig.~\ref{fig:E_s1_d3}), the gain drops to $\approx 6\%$, because for high $D_{max}$, the TXOP limit that can provide the necessary delay quantile becomes almost the same for the studied approaches.
Also, if we increase $\sigma$ to \SI{0.5}{\ms} (Fig.~\ref{fig:E_s5_d3}), the gain drops almost to $5\%$, because we need to occupy the channel for a longer time to compensate the high variance of RTA frame arrival time.	

For a high delay budget: $D_{max}=\SI{5}{\ms}$ (Fig.~\ref{fig:Q_s5_d5}), there is no need to decrease the TXOP limit.
As a result, the TXOP limit for two approaches becomes almost the same, and PCA does not provide better channel usage efficiency even for $T_{period}=\SI{100}{\ms}$ (Fig.~\ref{fig:E_s5_d5}).

To sum up, PCA is more efficient when very low delay is required or when the RTA frame periodicity is below a specific value.
The increase in the RTA frame arrival time variance reduces the gain from PCA and decreases the range of RTA frame periodicity for which this approach is profitable.

\section{Conclusion}\label{sec:outro}
We have studied two simple yet easy-to-implement approaches for RTA QoS provision.
Both approaches rely on special EDCA parameters for the RTA traffic to allow the RTA frames to be served without the contention with non-RTA traffic.
The first approach is to decrease the TXOP limit for non-RTA traffic so that the RTA traffic does not have to wait too long for the end of non-RTA transmissions.
The second approach is based on PCA with RTS/CTS.
To enable this approach, the 802.11be amendment shall allow a STA to generate an RTS without a present data frame.

We have developed mathematical models of both approaches that can be used to find the optimal configuration of these approaches.
We use the developed models to find the gain in channel usage efficiency from PCA and the RTA frame periodicity for which it is profitable to use this method.

We have found out that for RTA traffic with not too strict delay requirements or with very high intensity, it is better to decrease the TXOP limit, while PCA is better for strict delay limitations or not too frequent traffic.
At the same time, the high variance of RTA frame arrival time can decrease the efficiency of PCA.
		
In our future research, we plan to evaluate the efficiency of the considered approaches in scenarios with several RTA STAs and with overlapping BSSs.
	\bibliographystyle{IEEEtran}
	\bibliography{biblio.bib}
\end{document}